\documentclass{SCIS2021}
\usepackage{xcolor}
\UseRawInputEncoding

\begin{document}

\ArticleType{RESEARCH PAPER}
\Year{2021}
\Month{}
\Vol{}
\No{}
\DOI{}
\ArtNo{}
\ReceiveDate{}
\ReviseDate{}
\AcceptDate{}
\OnlineDate{}

\title{NAND-SPIN-Based Processing-in-MRAM Architecture for Convolutional Neural Network Acceleration}

\author[1,5]{Yinglin ZHAO}{}
\author[2]{Jianlei YANG}{{jianlei@buaa.edu.cn}}
\author[3]{Bing LI}{{bing.li@cnu.edu.cn}}
\author[2]{Xingzhou CHENG}{}
\author[2]{\\Xucheng YE}{}
\author[4]{Xueyan WANG}{}
\author[4]{Xiaotao JIA}{}
\author[4]{Zhaohao WANG}{}
\author[1]{\\Youguang ZHANG}{}
\author[4]{Weisheng ZHAO}{}

\AuthorMark{Zhao Y}



\address[1]{School of Electronic and Information Engineering, Beihang University, Beijing {\rm 100191}, China}
\address[2]{School of Computer Science and Engineering, Beihang University, Beijing {\rm 100191}, China}
\address[3]{Academy for Multidisciplinary Studies, Capital Normal University, Beijing {\rm 100048}, China}
\address[4]{School of Integrated Circuit Science and Engineering, Beihang University, Beijing {\rm 100191}, China}
\address[5]{Qingdao Research Institute, Beihang University, Qingdao {\rm 266104}, China}

\abstract{The performance and efficiency of running large-scale datasets on traditional computing systems exhibit critical bottlenecks due to the existing “power wall” and “memory wall” problems. To resolve those problems, processing-in-memory (PIM) architectures are developed to bring computation logic in or near memory to alleviate the bandwidth limitations during data transmission. NAND-like spintronics memory (NAND-SPIN) is one kind of promising magnetoresistive random-access memory (MRAM) with low write energy and high integration density, and it can be employed to perform efficient in-memory computation operations. In this work, we propose a NAND-SPIN-based PIM architecture for efficient convolutional neural network (CNN) acceleration. A straightforward data mapping scheme is exploited to improve the parallelism while reducing data movements. Benefiting from the excellent characteristics of NAND-SPIN and in-memory processing architecture, experimental results show that the proposed approach can achieve $\sim$2.6$\times$ speedup and $\sim$1.4$\times$ improvement in energy efficiency over state-of-the-art PIM solutions.}


\maketitle

\section{Introduction}

Over the past decades, the volume of data required to be processed has been dramatically increasing\cite{shafique2017adaptive}. As the conventional von Neumann architecture separates processing and data storage components, the memory/computational resources and their communication are in the face of limitations due to the long memory access latency and huge leakage power consumption. This phenomenon can be interpreted as memory and power walls \cite{luo2021spinlim}. Therefore, there is an urgent need to innovate the architecture and establish an energy-efficient and high-performance computing platform to break existing walls. 

Processing-in-memory (PIM), a promising architecture diagram, has been proposed to overcome power and memory walls in recent years \cite{cai2021proposal, liu2018processing}. Through the placement of logic units in the memory, the PIM architecture is considered an efficient computing platform because it performs logic operations by leveraging inherent data-processing parallelism and high internal bandwidth \cite{song2018graphr, eckert2018neural}. However, the full exploitation of the bandwidth and the integration of computing cells within the memory result in a major circuit redesign and a significant chip area increase \cite{hao2021recent}. As CMOS technology is moving to its physical limitation \cite{papandroulidakis2019practical}, the realization of PIM generates increases design and manufacturing costs and sacrificed memory capacity to some extent, which is not conducive to obtaining cost-effective products.

In recent years, many non-volatile memories (NVMs), such as resistive random-access memory (ReRAM) \cite{xue201924,li2018reram, yuan2021nas4rram}, phase change memory (PCM) \cite{ kim2020evolution, ambrogio2018equivalent}, and magnetoresistive random-access memory (MRAM) \cite{guo2021spintronics, apalkov2016magnetoresistive}, provide PIM with a new research platform. Among all emerging NVM technologies, MRAM has emerged as a promising high-performance candidate for the main memory due to its non-volatility, superior endurance, zero standby leakage, compatibility with the CMOS fabrication process and high integration density \cite{jain2017computing}.  In particular, spin-transfer torque MRAM (STT-MRAM) and spin-orbit torque MRAM (SOT-MRAM) are two advanced types of MRAM devices \cite{wang2018field}. However, the switching speed and energy consumption of STT-MRAM are limited by the intrinsic incubation delay, while SOT-MRAM exhibits a poor integration density because it contains two transistors in a standard bit cell \cite{cai2021sub}. In \cite{wang2018high,shi2021experimental}, an emerging spintronics-based magnetic memory, NAND-like spintronics memory (NAND-SPIN), was designed to overcome the shortcomings of STT-MRAM and SOT-MRAM and pave a new way to build a novel memory and PIM architecture.

Convolutional neural networks (CNNs) have received worldwide attention due to their potential of providing optimal solutions in various applications, including popular image recognition and language processing \cite{angizi2018imce}. As neural networks deepen, the high-performance computation of CNNs requires a high memory bandwidth, large memory capacity, and fast access speed, which are becoming harder to achieve in traditional architectures. Inspired by the high performance and impressive efficiency of PIM, researchers have attempted to implement in-memory CNN accelerators. For example, CMP-PIM involves a redesign of peripheral circuits to perform CNN acceleration in the SOT-MRAM-based memory \cite{angizi2018cmp}. STT-CiM \cite {jain2017computing} enables multiple word lines within an array to realize in-memory bit-line addition through the integration of logic units in sense amplifiers. However, their performance improvement brought about by PIM is offset by the shortcomings of the SOT/STT-MRAM mentioned above.

NAND-SPIN adopts a novel design that allocates one transistor for each magnetic tunnel junction (MTJ) and writes data with a small current, which means low write energy and high integration density. Despite its excellent potential, the PIM architecture based on NAND-SPIN is still scarce. In this study, we developed an energy-efficient memory architecture based on NAND-SPIN that can simultaneously work as an NVM and a high-performance CNN accelerator. The main contributions of this study are summarized as follows:

\begin{itemize}

\item Inspired by the outstanding features of NAND-SPIN devices, we developed a memory architecture based on NAND-SPIN. Through the modification of peripheral circuits, the memory subarray can perform basic convolution, addition and comparison operations in parallel.

\item By breaking CNN inference tasks into basic operations, the proposed NAND-SPIN-based PIM architecture achieves a high-performance CNN accelerator, which has the advantages of in-memory data movement and excellent access characteristics of NAND-SPIN.

\item We employed a straightforward data mapping scheme to fully exploit data locality and reduce data movements, thereby further improving the performance and energy efficiency of the accelerator. 

\item Through bottom-up evaluations, we show the performance and efficiency of our design with comparison to state-of-the-art in-memory CNN accelerators.
\end{itemize}

The remainder of this paper is organized as follows: Section 2 presents the background of MRAM and design motivation. Section 3 provides the details of the proposed architecture. Section 4 presents the acceleration methods for CNNs and introduces some optimization schemes. Section 5 describes the experimental platform and analyzes the simulation results. Section 6 concludes this paper.

\section{Preliminary and Motivation}

In this section, we present MRAM-related technologies, CNNs and existing in-memory computing designs.

\subsection{MRAM}
\begin{figure}[thbp]
\centering
\includegraphics[width=0.9\textwidth]{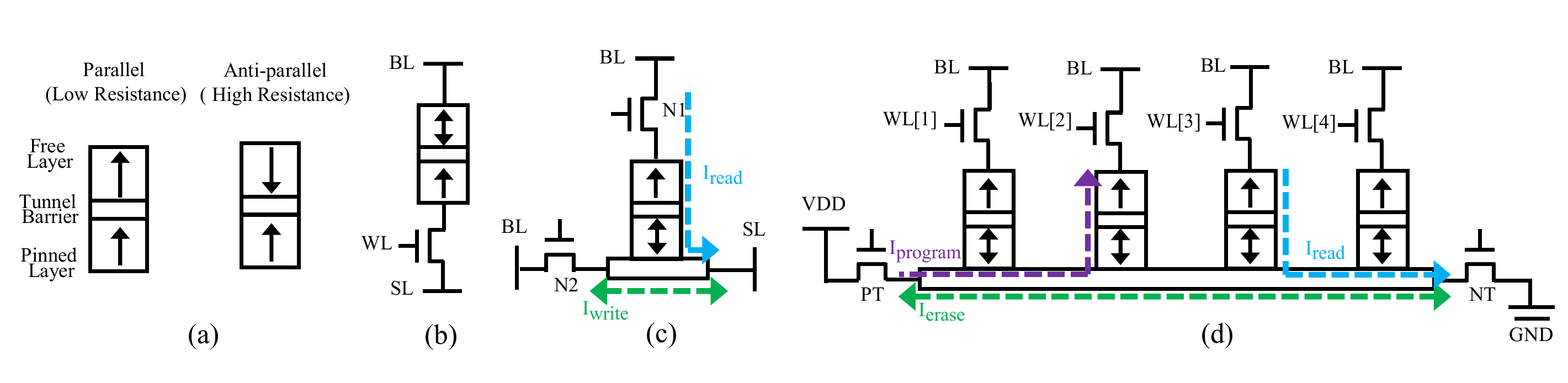}
\caption{ (a) Device structure of the MTJ in parallel and anti-parallel states. (b) 1T-1MTJ STT-MRAM cell. (c) Bit cell schematic of the standard 2-transistor SOT-MRAM. (d) Structure and operations of the NAND-like spintronic memory.}
\label{STT_SOT}
\end{figure}

MTJs are the basic storage element in STT-MRAM and SOT-MRAM \cite{wang2018field,Cai2021ASO}. As shown in Fig.~\ref{STT_SOT}a, an MTJ contains three layers: two ferromagnetic layers with a tunnel barrier sandwiched between them. The magnetization direction of the pinned layer is fixed and perpendicular to the substrate surface, while the magnetization direction of the free layer exhibits two stable states: parallel (P) or anti-parallel (AP) to that of the pinned layer. Due to the tunnel magnetoresistance (TMR) effect, when the magnetization directions of the two ferromagnetic layers are parallel (anti-parallel), the resistance of the MTJ is low (high). This state is used to represent the logic ``0'' (``1'') \cite{fong2016spin}.

The most popular STT-MRAM cell structure is illustrated in Fig.~\ref{STT_SOT}b \cite{rho201723}. The MTJ pillar has a small area and can be integrated above transistors. Hence, the total cell area is determined only by the bottom transistors and leads to an expectation of achieving a high-density memory. However, the long write latency and high write energy hinder the broad application of STT-MRAM.

SOT-MRAM is a composite device of spin hall metal and MTJ \cite{guo2021spintronics}, and Fig.~\ref{STT_SOT}c shows the basic bit cell of a standard SOT-MRAM. The access transistors, N1 and N2, connect the pinned layer of the MTJ and heavy metal strip with bit lines (BLs), respectively. The data can be written into and read out from the MTJ by referring to the green and blue currents flowing from the source lines (SLs) to BLs \cite{peng2020exchange}. Although SOT brings the fast switching of magnetization, such a design faces the storage density challenge because it contains two transistors in a unit.

A multi-bit NAND-SPIN device is shown in Fig.~\ref{STT_SOT}d, in which the MTJs are organized similar to a NAND flash memory \cite{wang2018high,yu2021proposal}. The PMOS transistor (PT) and NMOS transistor (NT) work as the selection transistors for conducting paths to the VDD and GND, respectively. In the NAND-SPIN, the write operation requires two steps: 

\textbf{Step 1}: Erase data in all MTJs, and initialize them into default AP states. In this step, two transistors, PT and NT, are activated, while all word line (WL) transistors are off. The generated current between VDD and GND can erase all MTJs in the heavy metal strip via the SOT mechanism.

\textbf{Step 2}: Program the selected MTJs by switching them into the P state. In this step, the corresponding WL and PT transistors are activated, and the currents flowing through the MTJs from free layers to pinned layers would switch the states of the MTJs to the P state via the STT mechanism. 

Because NAND-SPIN uses MTJs as the basic storage element, it has high endurance, which is essential for memory cells. In addition, the compatibility with CMOS makes NAND-SPIN a high density memory, because it distributes MTJs over CMOS circuits. Compared with conventional STT-MRAM, NAND-SPIN only requires a small STT current to complete an efficient AP-to-P switching. The asymmetric writing scheme reduces the average latency and energy of write operations while achieving a high storage density, which unlocks the potential of MRAM-based architectures.

\subsection{CNN}

A CNN is a type of deep neural network, commonly used for image classification and object recognition. Typically, a CNN consists of three main types of layers, namely, convolutional layer, pooling layer and fully-connected layer \cite{eckert2018neural, 7551379, yang2021s2engine}. 

In the convolutional layer, the kernels extract features from the input feature maps through convolution operations. The convolution operation applies a kernel to move across the input feature map and performs dot products between the inputs and weights. There are usually many input and output feature maps in a convolutional layer, which requires considerable convolution operations. 

The pooling layer is used to reduce the input dimensions of the feature maps. Similar to the convolutional layer, the pooling operation slides a filter across the inputs and combines the neuron clusters into a single neuron. There are two types of pooling layers, namely max/min pooling and average pooling. Max/min pooling uses the maximum/minimum value of each cluster as the neuron of the next layer, while average pooling uses the average value.

The fully-connected layer connects all neurons from one layer to every activation neuron of the next layer, and it usually leverages a softmax activation function to classify inputs as the final outputs. Past studies have concluded that the fully-connected layer can be treated as another convolutional layer \cite{zhou2016dorefa, angizi2019mrima}.

\subsection{PIM Architectures}

To reduce the cost of data movement, the PIM platform has been proposed for several decades \cite{ghose2019processing, imani2019floatpim, wang2021triangle}. Some proposals in the context of static RAM (SRAM) or dynamic RAM (DRAM) have been researched in recent years. For example, in \cite{chen2017Eyeriss}, a grid of SRAM-based processing elements was utilized to perform matrix-vector multiplication in parallel. The design in \cite{li2017drisa} uses a CNN accelerator built with DRAM technology to provide a powerful computing capability and large memory capacity. However, their working mechanisms inevitably lead to multi-cycle logic operations and high leakage power.

Considering the possibility of using NVM as a substitute for the main memory, various works have been conducted to explore emerging PIM architectures. These works put forward a wide range of specialized operators based on NVM for various applications \cite{wang2020tcim, yang2015radiation}. For instance, in \cite{cai2022stateful}, an interesting design was proposed to implement in-memory logic based on MTJs. Pinatubo optimized the read circuitry to perform bitwise operations in data-intensive applications \cite{li2016pinatubo}. Based on PCM, a equivalent-accuracy accelerator for neural network training is achieved in \cite{ambrogio2018equivalent}. In addition, some designs modify memory peripherals to perform specific applications instead of general applications. In \cite{tang2017binary}, a ReRAM crossbar-based accelerator was proposed for the binary CNN forward process. Moreover, PRIME shows a ReRAM-based PIM architecture in which a portion of a memory array can be configured as NN accelerators \cite{chi2016prime}.

Although PIM-based designs effectively reduce data movements, the complex multi-cycle operations and insufficient data reuse are still hindrances to performance improvement. Different from previous designs, we not only used NAND-SPIN to build an in-memory processing platform, but optimized the storage scheme to minimize data duplication and provide large parallelism for in-memory processing. 
\begin{figure}[t]
\centering
\includegraphics[width=0.6\textwidth]{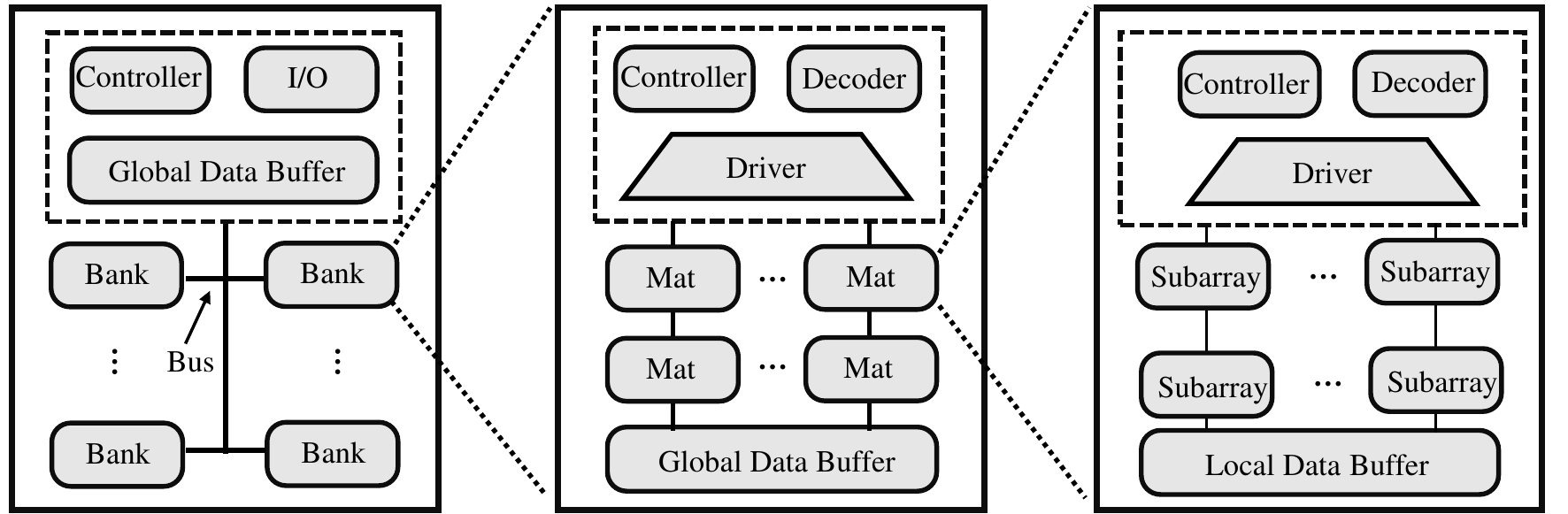}
\caption{Hierarchical memory organization in the proposed architecture.}
\label{top_arch}
\end{figure}
\section{Proposed Architecture}
 
In this section, we first introduce the architecture design and the function of each component. Then, we show how to perform memory and logic functions based on the proposed architecture. 

\subsection{Architecture}

\begin{figure}[t]
\centering
\includegraphics[width=0.9\textwidth]{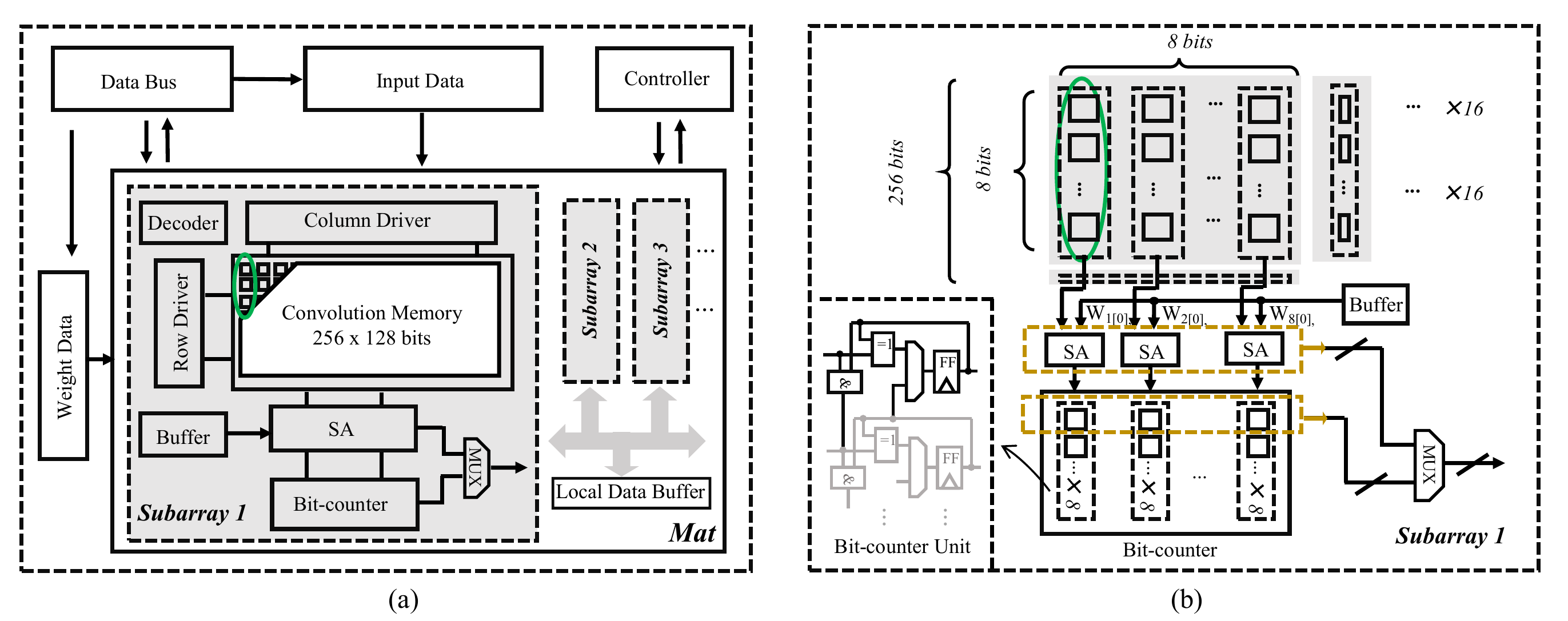}
\caption{(a) Mat level scheme and peripheral components.  (b) NAND-SPIN-based subarray architecture.}
\label{arch}
\end{figure}

The general memory organization is shown in Fig.~\ref{top_arch}. There are three levels in such a hierarchical organization: the bank, mat and subarray. The bank is a fully-functional memory unit and banks within the same chip share the I/O resources. The mat is the building block of bank, and multiple mats are connected with a global data buffer. The subarray is the elementary structure in our design, and multiple subarrays in a mat implement memory access or CNN acceleration in parallel. To coordinate those components, the controller generates control signals to schedule computations and communications. In particular, the local data buffer temporarily hold data sent from subarrays and the global buffer for alleviating data congestion. The mat level scheme and peripheral components is shown in Fig.~\ref{arch}a, and the subarray architecture based on NAND-SPIN is illustrated in Fig.~\ref{arch}b.  Here, we mark a single NAND-SPIN device containing a group of 8 MTJs with a green ellipse. The specific structure of subarrays and the operation details of CNN acceleration are discussed later.

\subsection{Microarchitecture}

\begin{figure*}[t]
\centering
\begin{minipage}[t]{0.49\textwidth}
\setcaptionwidth{2.9in}
\centering
\includegraphics[width=1.0\textwidth]{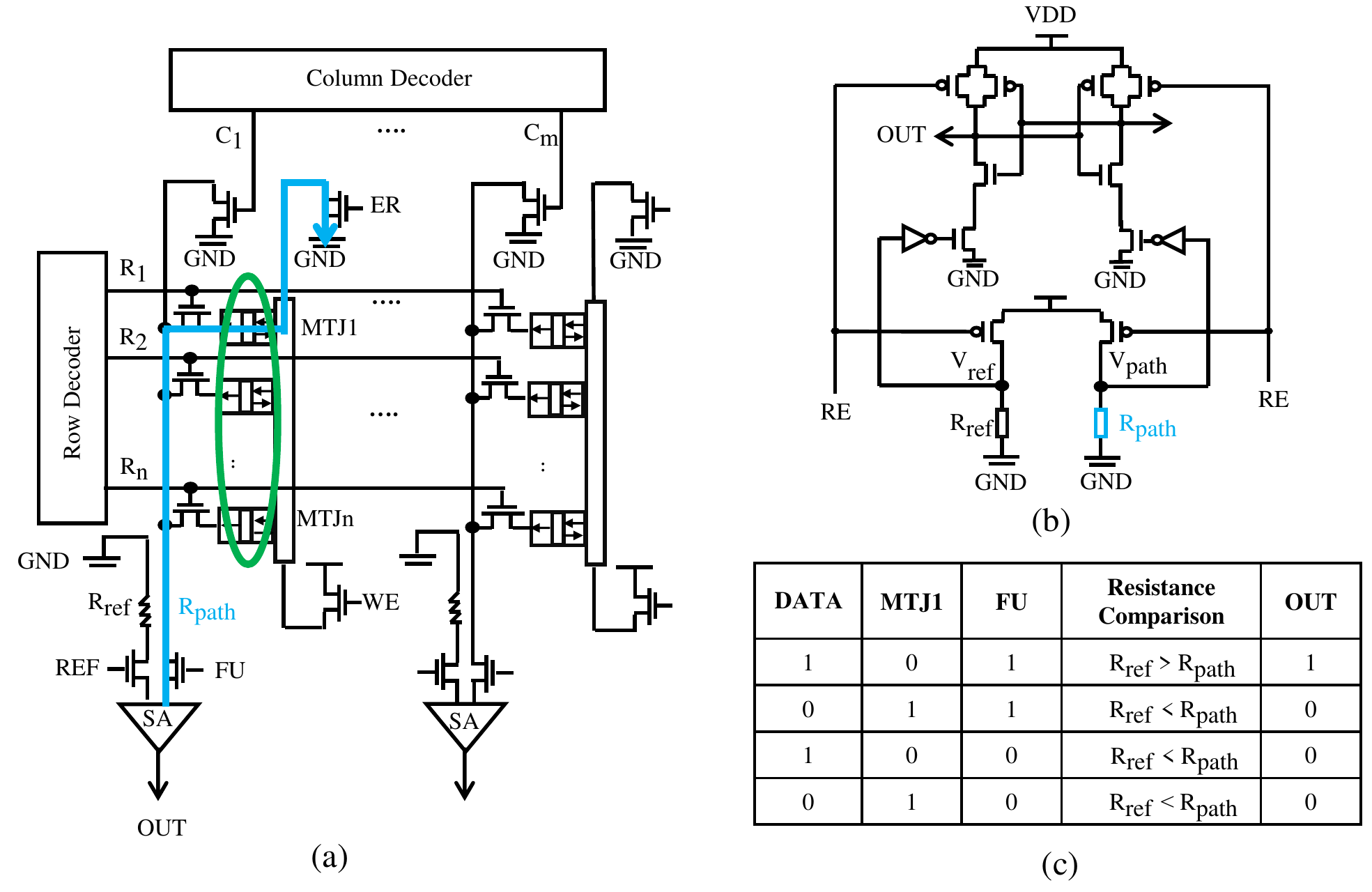}
\caption{(a) Schematic of the subarray architecture. (b) Schematic of the sensing circuit. (c) Possible conditions and outputs of the SA.}
\label{sub_group}
\end{minipage}
\begin{minipage}[t]{0.49\textwidth}
\setcaptionwidth{2.9in}
\centering
\includegraphics[width=1.0\textwidth]{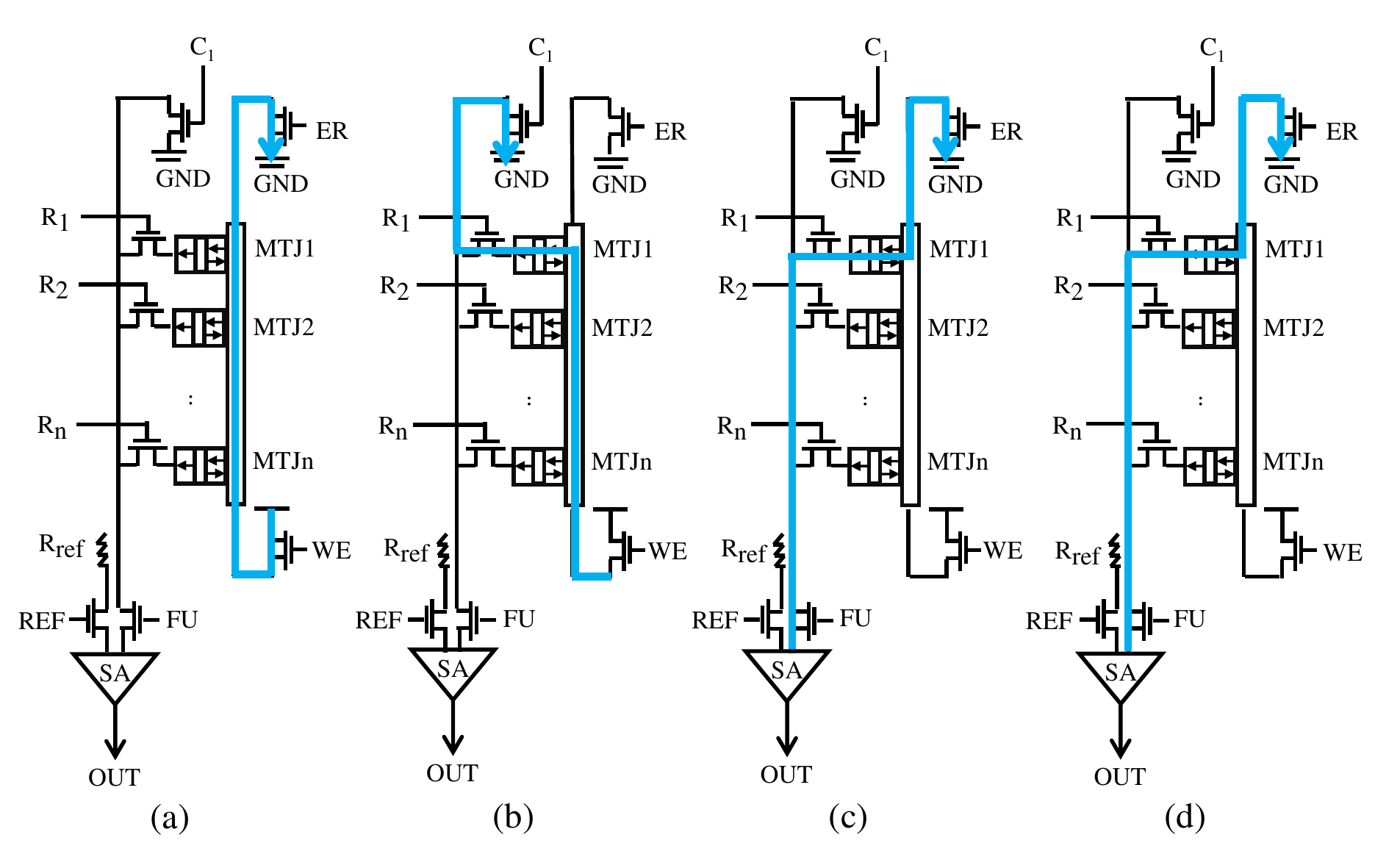}
\caption{ The current paths for (a) erase operation, (b) program operation, (c) read operation, (d) AND operation.}
\label{w_r_p}
\end{minipage}
\end{figure*}

Fig.~\ref{sub_group}a describes the detailed structure and internal circuits of a block. The synergy of control signals carries out 3 logic functions: writing, reading and logic AND (for CNN acceleration mode). The writing process is divided into two stages: the stripe erase stage and the program stage. As illustrated in Section 2.1, the WE and ER are both activated in the erase stage to generate the SOT current, while the WE, C$_{x} (x = [1, m])$ and corresponding R$_{y} (y = [1, n])$ are selected in the later program stage to produce the STT current. In regard to read operations, the REF, FU and corresponding R$_{y} (y = [1, n])$ are set high. Then, the SA is connected to the circuit for a reading operation. Besides, the setting for AND operations is similar to read operations, but the FU varies with the operand.

The SA is the central functional unit that performs read operations and AND operations, utilizing a separated PCSA (SPCSA) circuit (depicted in Fig.~\ref{sub_group}b) \cite{zhang2017reliability}. The SPCSA can sense the resistance difference between two discharge branches according to the discharge speed at two points (V$_{ref}$ and V$_{path}$). Accordingly, R$_{ref}$ refers to the resistance in the reference path, and is set to (R$_{H}$+R$_{L}$)/2 (R$_{H}$ and R$_{L}$ represent the resistance of an MTJ in AP and P states, respectively), and R$_{path}$ represents the total resistance in another path. 

An SA requires two steps to implement a single function. The first step is to charge V$_{ref}$ and V$_{path}$ by setting the RE low voltage. The second step is a reverse process that flips RE to discharge V$_{ref}$ and V$_{path}$. The inverter connected to the point with a higher path resistance first flips and latches the state. 

Note that we use a complementary method for data storage. For example, the MTJ in the AP state actually means storing binary data ``0''. Fig.~\ref{sub_group}c lists the possible conditions (DATA represents the actual binary data stored in MTJ1) and the outputs of the SA. Moreover, the transistor connected to the REF is turned on by default when the SA is working.

\textbf{1). Memory Mode}: Based on the subarray design described above, Fig.~\ref{w_r_p} and Table \ref{step} describe the paths of the current flow and corresponding signal states respectively.

\textbf{Erase operation}: To erase the contents in a group of MTJs, the current is generated flowing through the heavy metal strip. As shown in Fig.~\ref{w_r_p}a, the transistors in contact with heavy metal strips are activated by ER and WE, while the other transistors remain deactivated. Then, a path is formed between VDD and GND, and it generates the write current in the heavy metal strip to erase the MTJs to AP states.

\textbf{Program operation}: The program operation is the second step of data writing after the erase operation. A program operation requires a current from the free layer to the fixed layer in the MTJ, as shown in Fig.~\ref{w_r_p}b. While programming data (represented as D in Table \ref{step}), the circuit should activate the transistor controlled by WE and the two transistors corresponding to a certain MTJ (for example, R$_{1}$ and C$_{1}$ for MTJ1 in Fig.~\ref{w_r_p}b). Then, a path is formed between VDD and GND, which produces a current inducing the STT to switch the MTJ from AP to P. 

\begin{figure*}[t]
\centering
\begin{minipage}[t]{0.49\textwidth}
\setcaptionwidth{2.9in}
\centering
\includegraphics[width=0.8\textwidth, height=0.6\textwidth]{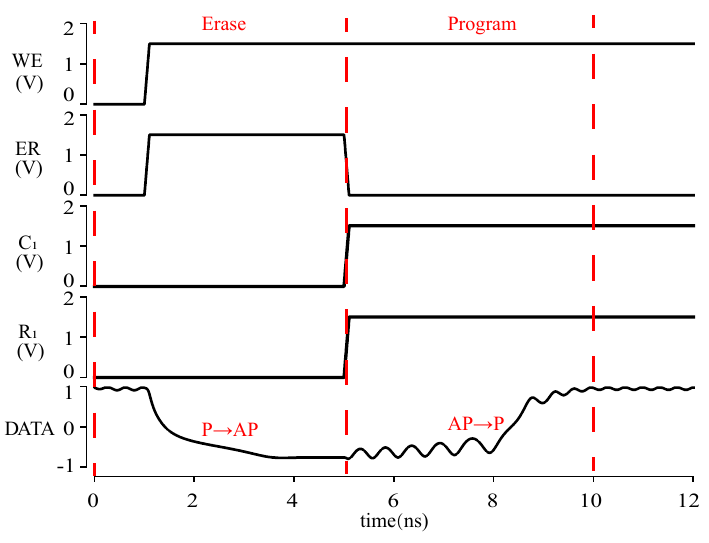}
\caption{ Timing diagram of erase and program operations.}
\label{erase_program}
\end{minipage}
\begin{minipage}[t]{0.49\textwidth}
\setcaptionwidth{2.9in}
\centering
\includegraphics[width=0.8\textwidth, height=0.6\textwidth]{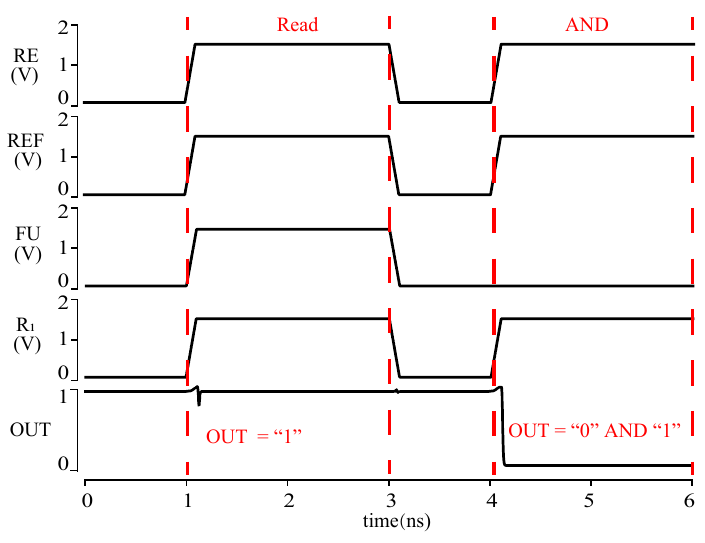}
\caption{ Timing diagram of read and AND operations.}
\label{read_and}
\end{minipage}
\end{figure*}

Note that the state of an MTJ after finishing the two stages above is determined by the signals sent from decoders. The signals (R$_{1}$ to R$_{n}$) determine which row performs the program operation. The signals (C$_{1}$ to C$_{m}$) produced by the column decoder determine whether the program operation is completed. Noticing the mapping relationship above, we regard generated signals as a map to values that need to be written into MTJs. The signal C$_{x} (x = [1, m])$ equal to ``1'' results in a successful program operation as well as the AP-to-P switching in the MTJ. In contrast, the logic “0” in C$_{x} (x = [1, m])$ means a blocking current in the transistor connected with C$_{x} (x = [1, m])$, and the MTJ maintains the AP state. Fig.~\ref{erase_program} demonstrates the timing diagram of an erase operation followed by a program operation.

\textbf{Read operation}: When performing a typical read operation, a current should be generated in the path connecting the SA and a certain MTJ, as shown in Fig.~\ref{w_r_p}c. Similar to the program operation, the signals (R$_{1}$ to R$_{n}$) transmitted by row decoders decide which row of MTJs would be read out. Additionally, ER, FU and REF need to be set to logic ``1'' during read operations, and then the states of  MTJs can be indicated by outputs of SAs. An output “0” indicates that the MTJ has a high resistance (AP state) and stores logic ``0''. Conversely, an output ``1'' refers to an MTJ storing ``1'' in the P state.

As our subarray structure is different from traditional architectures, the memory access scheme needs to be modified accordingly. In our design, the erase operation can reset a group of MTJs in a single NAND-SPIN device and is always followed by a set of program operations for writing data. However, a read operation does not involve other operations, which causes asymmetry in the read and write operations. In other words, the subarray writes a row of NAND-SPIN devices with an erase operation and $N$ program operations ($M\times N$ bits in total, where $M$ is the number of columns, $N$ is the number of MTJs in a NAND-SPIN device, and $M\times N$ is $128\times8$ in our design) instead of writing a row of MTJs with a single write operation like the traditional architecture \cite{angizi2019mrima}. Nevertheless, the read operation reads a row of data out (128 bits in our design) at a time, the same as the traditional architecture. 

\begin{table}[t]
\footnotesize
\caption{Circuit signals for different operations}
\label{step}
\tabcolsep 12pt 
\begin{tabular*}{\textwidth}{cccccccccc}
\toprule
  Operation&WE&ER&\(\rm{C_{1}}\) & \(\rm{R_{1}}\) & FU & REF & MTJ\footnote & MTJ\footnote  & OUT \\\hline
  Erase&1&1&0&0&0&0&/&1&/\\
  {Program D}&1&0&D&1&0&0&1&$\rm \overline{D}$&/\\
  Read&0&1&0&1&1&1&$\rm \overline{D}$&$\rm \overline{D}$&D\\
  AND&0&1&0&1&$\rm {W}$&1&$\rm \overline{D}$&$\rm \overline{D}$&{$\rm {W}$\ `AND'\ $\rm {D}$}\\
\bottomrule
\end{tabular*}
\end{table}
\footnotetext[1]{The MTJ state before the operation}
\footnotetext[2]{The MTJ state after the operation}

Due to the introduction of an erase operation before program operations, the write operation latency would be increased. However, the SOT-induced erase operation could reset multiple MTJs on the same heavy metal strip, while the program operations set MTJs individually. Therefore, the time consumed by a erase operation is amortized. In addition, the SOT-induced erase operation is much faster than the program operation induced by STT, which further offsets the extra latency.

It should be noticed that the read disturb could be significantly mitigated in our design. As the P-to-AP switching is induced by SOT and the AP-to-P switching is based on STT, the read disturb margin is related to the read current and the P-to-AP STT switching current. Therefore, we can increase the P-to-AP STT switching current of MTJs by adjusting the HM dimension to mitigate read disturb issues and enhance the reliability.

\textbf{2). CNN Acceleration Mode}: In CNN acceleration mode, the AND logic is activated in SAs. As shown in Fig.~\ref{w_r_p}d, the AND operation has the same current path as the read operation, and the difference between them lies in FU. FU is always at a high voltage during a read operation, while FU is used to represent one of the two source operands (represented as W in Table \ref{step}) during an AND operation. Another source operand is supposed to have been stored in the selected MTJ, and the SA finally obtains the AND operation result. Only when the MTJ is in a low resistance state (storing ``1''), FU is under high voltage (indicating logic ``1''), and the resistance of R$_{path}$ is smaller than R$_{ref}$, the SA outputs ``1''. Other situations result in R$_{path}$ being larger than R$_{ref}$, and the SA outputs ``0''. Fig.~\ref{read_and} demonstrates the timing diagram of a read operation and an AND operation, assuming that D = ``1'' and W = ``0''.

While accelerating CNN inferences, data buses are used for transmission of weight and input data, both of which are considered as collections of source operands (especially for AND operations). The weight and input data need to be transferred into the buffers and convolution memories (CMs) before the CNN computation starts. The buffer is used for storing temporary weight data to reduce data movements and bus occupation. Moreover, the buffer is connected to the data bus through private data ports so that it does not occupy the bandwidth of the subarray. The bit-counter in each column could count the non-zero values of all AND operation results obtained in the corresponding SA. The multiplexers are used to output the data sensed in SAs during normal read operations or the data in the bit-counters (bit-by-bit for each unit) during convolution operations, as shown in Fig.~\ref{arch}.

\section{Implementation}

This section first introduces the complex computing primitives in CNN computation, and then shows how our architecture performs an inference task. As introduced above, the convolutional layer involves considerable convolution operations, and the pooling layer performs iterative addition, multiplication and comparison operations to implement average pooling or max/min pooling. Since AND is a universal logic gate, we use it to implement computing primitives together with bit-counters.

\subsection{Building Blocks of CNN}

\textbf{Convolution}: Convolution is the core operation of CNN, and it takes up most fraction of computation resources. We consider $I$ ($W$) as an input (weight) fixed-point integer sequence located in an input (kernel) map \cite{zhou2016dorefa}. Assuming that $I=\sum_{n=0}^{N-1} c_{n}(I) 2^{n}$ and $W=\sum_{m=0}^{M-1} c_{m}(W) 2^{m}$ where $ ({c_{n}(I)})_{n=0}^{N-1}$ and $({c_{m}(W)})_{m=0}^{M-1}$ are bit vectors, the dot product of $I$ and $W$ can be specified in Eq.~\ref{con}.

\begin{small}
\begin{equation}
I\ast W = \sum_{n=0}^{N-1} \sum_{m=0}^{M-1} 2^{n+m} bitcount(AND(c_{n}(I),c_{m}(W))) .
\label{con}
\end{equation}
\end{small}

Regarding the computationally expensive convolution operation as a combination of rapid and parallel logic AND, bit-count and shift operations, the PIM architecture commonly converts it into consecutive bitwise operations. Previously, some schemes first store the weight and input data in the same column, and then sense the bitwise operation outputs in modified circuits \cite{angizi2019mrima, jain2017computing}. However, those methods require additional data duplication and reorganization while the weight matrix slides, which aggravate the overhead as the time-consuming and power-consuming characteristics of the NVM. 

\begin{figure*}[t]
\centering
\includegraphics[width=1.0\textwidth]{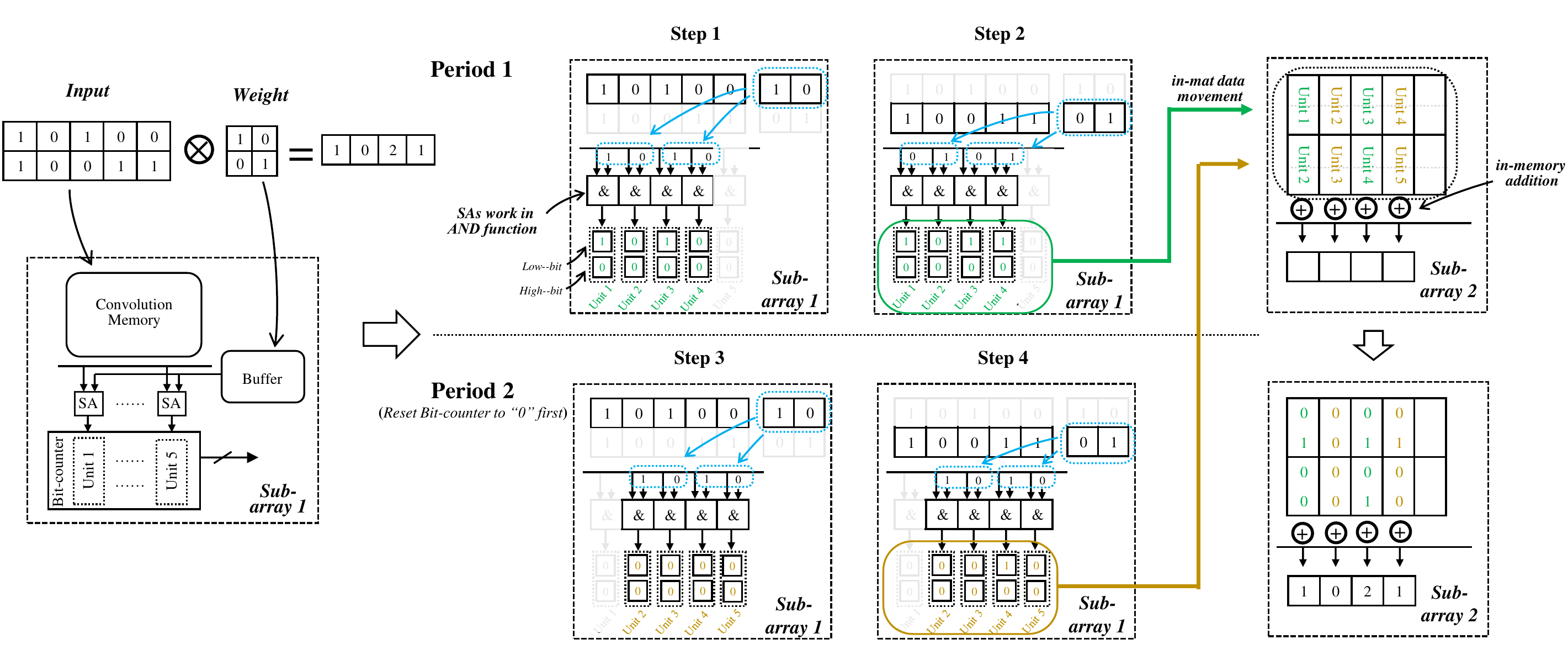}
\caption{Bitwise convolution operation.}
\label{op_detail}
\end{figure*}

To address this issue, we adopt a straightforward data storage scheme to reduce redundant access operations. We split both the input and weight data into 1-bit data. For example, an $M$-bit input matrix is converted to $M$ 1-bit matrices and stored in $M$ subarrays, and an $N$-bit weight matrix is decomposed into $N$ 1-bit matrices and transmitted to each subarray for bitwise convolution. Fig.~\ref{op_detail} illustrates the bitwise convolution of a 2$\times$2 weight matrix and a 2$\times$5 input matrix. In the first step, the first row of the input matrix in CM is activated, and the first row of the weight matrix in the buffer is connected to SAs in parallel for AND operations. The results are transferred to and counted in the bit-counter unit of each column. By repeating the above processes for the second row of matrices, the second step obtains the counting results in bit-counter units. Those units transfer their contents to Subarray 2 through in-mat data movement, and they would be reset to zero at the end of first period. The second period slides the weight matrix to the next position to work out another set of bit-counting results. Finally, Subarray 2 perform in-memory addition (will be discussed later) to get the bitwise convolution results.

Note that our design improves parallelism by greatly reusing the weights instead of duplicating the inputs in subarrays. In addition, the introduction of the buffer reduces the overhead of in-memory data movement. Requiring only one writing operation into the buffer, the 1-bit weight matrix would be used during the bitwise convolution operations of the entire 1-bit input matrix in this subarray, which significantly reduces data movements and dependence on the data bus. Since the buffer only needs to hold one bit of each weight matrix element, it does not require much capacity. 

\begin{figure*}[tbp]
\centering
\begin{minipage}[t]{0.49\textwidth}
\setcaptionwidth{3in}
\centering
\includegraphics[width=1.0\textwidth]{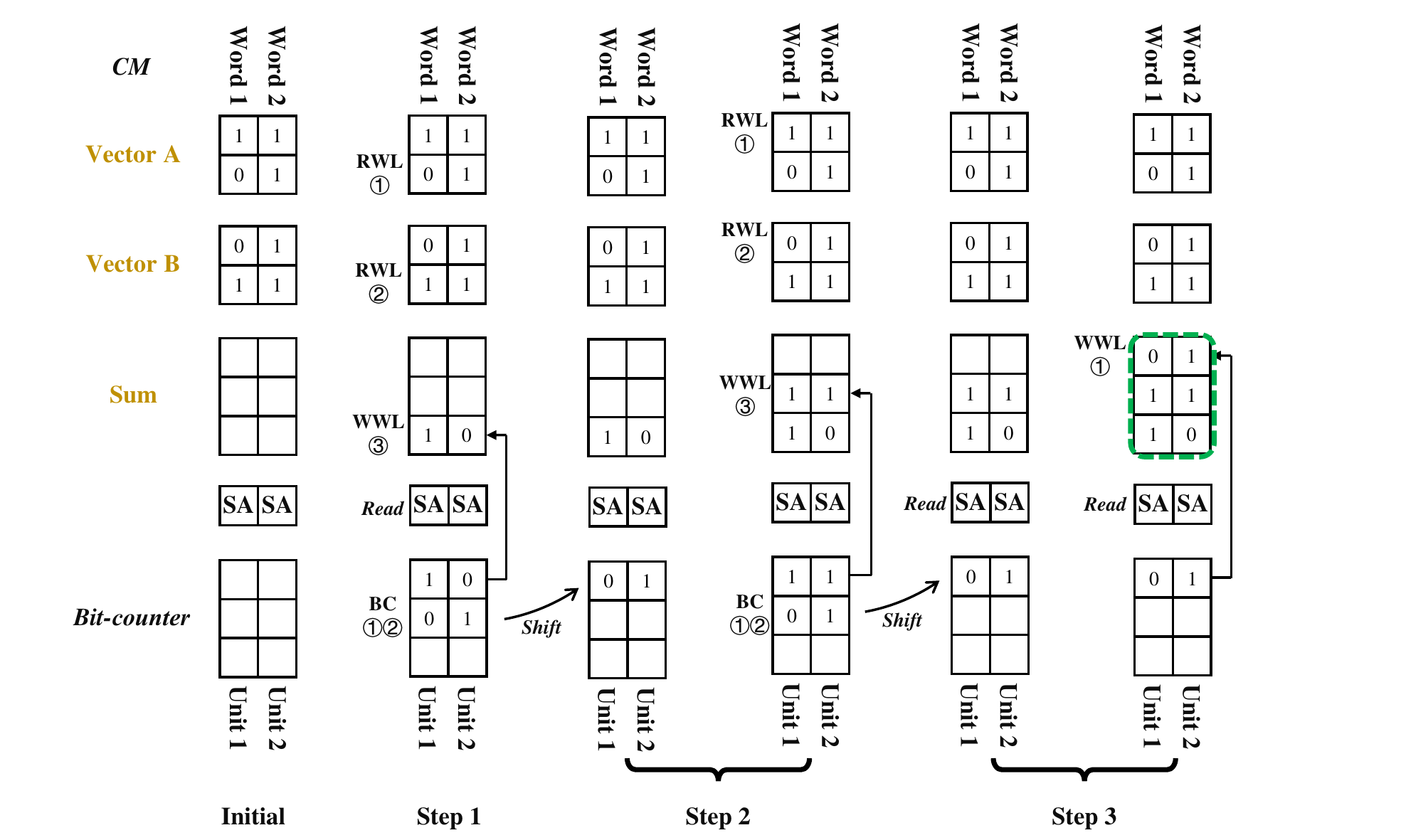}
\caption{Computation steps of the addition operation.}
\label{add}
\end{minipage}
\begin{minipage}[t]{0.49\textwidth}
\setcaptionwidth{3in}
\centering
\includegraphics[width=1.0\textwidth]{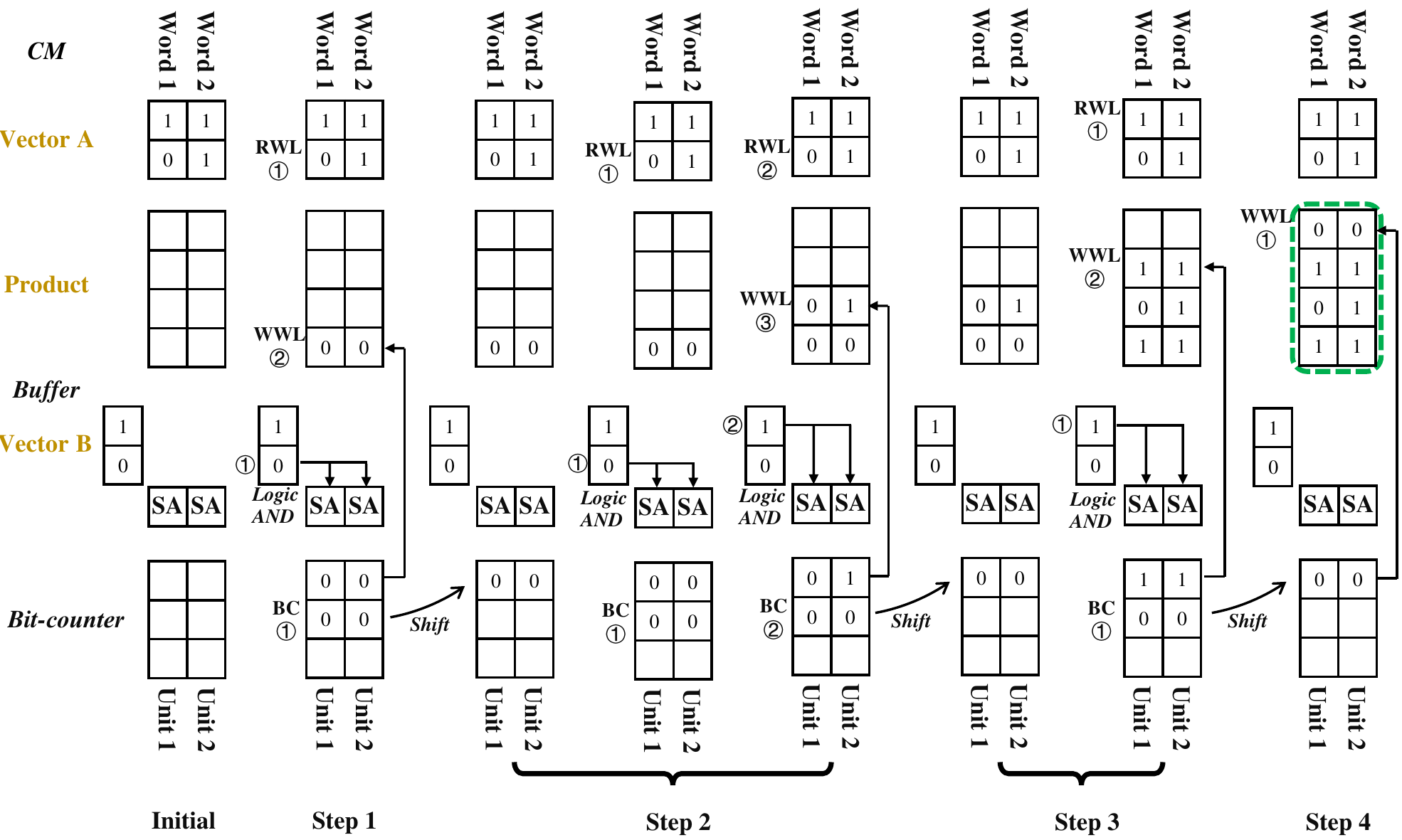}
\caption{Computation steps of the multiplication operation.}
\label{mult}
\end{minipage}
\end{figure*}

\textbf{Addition}: Unlike convolution, addition employs a data allocation mechanism that stores data element-by-element vertically \cite{eckert2018neural}. Before addition starts, all bits of the data elements are transposed and stored in the CM. One type of conventional design paradigm generally selects two rows of data simultaneously and performs addition operation using a modified sense amplifier. However, the process variation may cause logic failures, making it hard to guarantee reliability. Our design uses bit-counters to count the non-zero data in each bit-position from the least significant bit (LSB) to the most significant bit (MSB). Fig.~\ref{add} shows the data organization and addition work steps of two vectors (vector A and B, both are 2-bit numbers). The numbers in circles indicate the execution order of the involved operations in each step. The two vectors that are going to be added together are put in the same column of the CM. There are 3 empty rows reserved for the sum results. In each step, the bits of the two vectors at the same bit-position are read out by read WLs (RWL) and bit-countered (BC) in bit-counter units. The LSBs of the count results are written back through a write WL (WWL), and the other bits of the count results are right-shifted as the initial state of the next step. As demonstrated in Fig.~\ref{add}, the LSBs of the count results generated in the second and third steps are stored back as the second and third bits of the sum results. Moreover, the addition operation can be extended to the case where multiple source operands are added, as long as these operands are in the same column.

\begin{figure*}[t]
\centering
\includegraphics[width=1.0\textwidth]{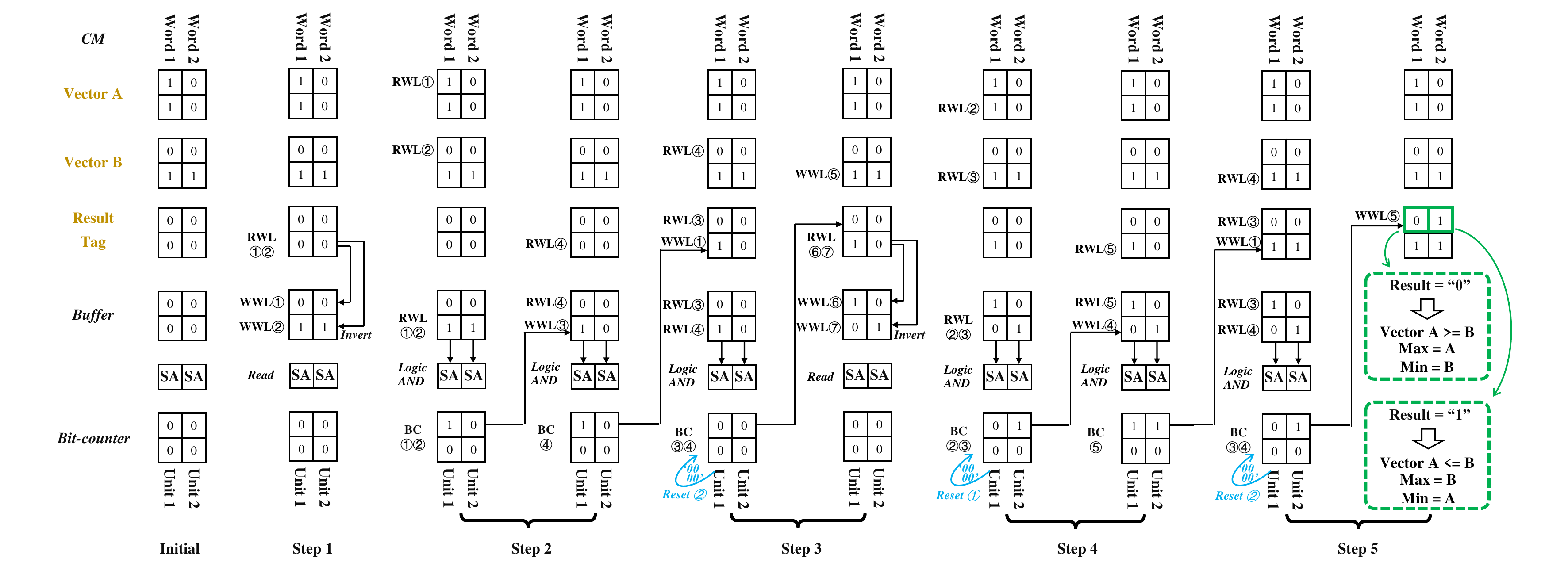}
\caption{Execution steps of the comparison operation.}
\label{comp_fig}
\end{figure*}

\textbf{Multiplication}: Multiplication has a data allocation mechanism similar to addition. The difference between them lies in that the AND function is activated in SAs to generate bit multiplication results. We show how multiplication works using an example of a 2-bit multiplication in Fig.~\ref{mult}. The multiplication starts with initializing all bits of two vectors (A and B) in the CM and the buffer, and there are 4 empty rows reserved for the product results. The multiplication algorithm generates the product results bit-by-bit from the LSB to the MSB. In each step, each bit of the product is produced by bit-counting all the single-bit products that corresponding to this bit-position. For example, since the LSBs of the products are determined by the bit multiplication results of the LSBs of two vectors (A and B), the LSBs of two vectors A and B are read out simultaneously to perform bit multiplication in the first step. Considering two bits read out as operands, the SAs perform parallel AND operations and transfer the results to bit-counter units for counting. Then, the LSBs of those units report the LSBs of the product and are stored back in CM (product part) accordingly by a WWL operation. The other bits of the count results, which record the carry-in information, are right-shifted as the initial state of the next step. Obviously, the second step requires more cycles to count two partial AND operation results than the first step. It should be noted that the buffer capacity is limited, so it is not wise to set a different multiplier for the multiplicand in each column. Therefore, our architecture is suitable for multiplicative scaling with the same scale factor.

\textbf{Comparison}: Max/Min extraction is a common operation in the max/min pooling layer. We demonstrate how to compare two sets of data (vector A and B) and select the max/min using the method shown in Fig.~\ref{comp_fig}. Initially, two vectors are stored bit-by-bit in the vertical direction along the BL. In addition, two extra rows of storage (Result and Tag) are both reset to 0, where Result row indicates the comparison results and Tag row is used as identifiers. In the first step, the row of Tag is read out by an RWL, and then two WWLs are activated to write the Tag row and its inverted values into the buffer. The second step activates two RWLs to read out the MSBs of the two vectors (A and B) on the same BL, and the SAs simultaneously perform AND operations considering the second row of the buffer as another operand. The outputs of SAs are subsequently bit-counted in the bit-counter. Then the LSB of each unit indicates the comparison result of two vectors. The LSB of the unit equaling 1 means that the two bits read out are different. Subsequently, we write the LSBs into the second row of the buffer and update the  bit-counter with the 'AND' operation results between the first row of the buffer and the Tag row. Next, the LSBs of bit-count units are written into the Tag row, and all bit-counter units are reset to zero. In step 3, as shown in Fig.~\ref{comp_fig}, two more AND operations are performed, where the MSBs (vector B), the Result row and the buffer are considered as operands. So far, the LSBs of bit-count units represent the comparison results only considering the first bit of each vector. We store the results in the Result row and start the next bit comparison process. The data in the Result and Tag rows are gradually updated as each bit is compared from MSB to LSB. If the final data located in the Result row is 1, vector A is greater than or equals to vector B, and A/B stands for the max/min of the two. Conversely, the binary data 0 means that B/A is the max/min.

\begin{figure}[t]
\centering
\includegraphics[width=1.0\textwidth]{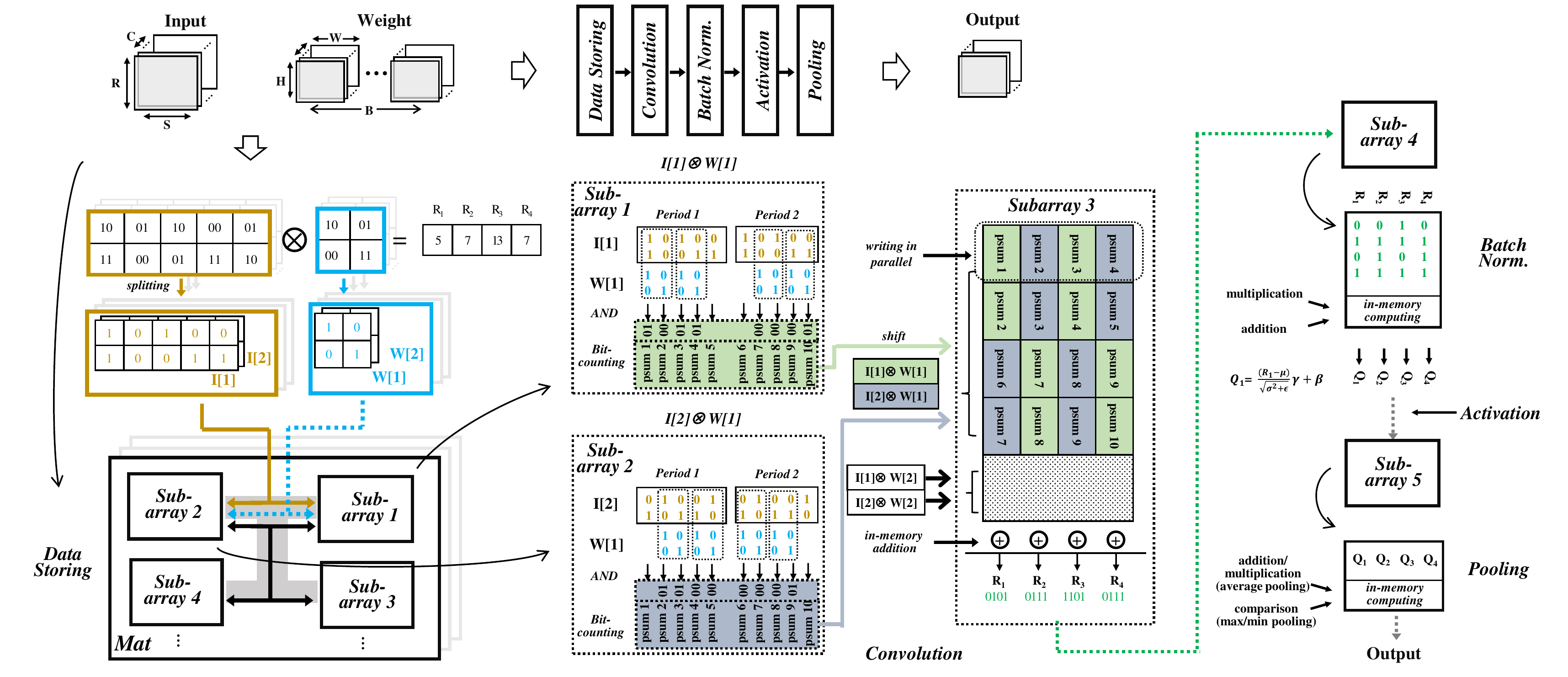}
\caption{Data organization and computation steps of CNN.}
\label{CNN_comp_step}
\end{figure}

\subsection{CNN Inference Accelerator}

In realistic scenarios of mainstream CNNs, it is hard to store all the data of one layer in a limited-capacity PIM platform. Therefore, reducing data duplication enables the memory array to accommodate more data. Fig.~\ref{CNN_comp_step} shows the data organization and computation steps of CNNs. Initially, the input matrix is split and organized in different subarrays in a mat. To perform CNN inference tasks, the weight matrix is decomposed and transferred into multiple subarrays for parallel bitwise convolution. Although there is still massive necessary data movements, our design tends to exploit the internal data buses, which can reduce the dependence on the external buses. The operations of each layer are described below.

\textbf{Convolutional layer}: In this layer, the subarrays are configured to generate partial-sums through bitwise convolution operations. The partial-sums are summed, and then sent to the activation function. To maximize parallelism, we adopt a cross-writing scheme during convolution operations. This scheme guarantees that the bit-counting results produced by different subarrays during the same period are not crossed. For example, as shown in Fig.~\ref{CNN_comp_step}, during the Period 1, Subarray 1 and 2 obtain the bit-counting results, which are not crossed and therefore could be written into different columns of the Subarray 3. Thus, the partial-sums are written in parallel without cache operations. In addition, since the bit-counting results are read out bit-by-bit from LSBs to MSBs, the shift operation can be realized by simply writing them to different rows in the vertical direction in Subarray 3.

In CNN, calculations with high-precision numerical values require significant computational power and storage resources. Quantization is the transformation process of lessening the number of bits needed to represent information, and it is typically adopted to reduce the amount of computation and bandwidth requirement without incurring a significant loss of accuracy. Several works have shown that the quantization to 8-bit can achieve comparable prediction accuracy as 32-bit precision counterparts \cite{zhou2016dorefa, colangelo2018exploration}. In our design, we perform the quantization using the minimum and the maximum values of the given layer. The transformation, which quantizes the input $Q_{i}$ to a $k$-bit number output $Q_{o}$, is as follows:

\begin{small}
\begin{equation}
Q_{o} = round((Q_{i}-Q_{min})\frac{(2^{k}-1)}{Q_{max}-Q_{min}}) .
\label{quantization}
\end{equation}
\end{small}
 
\noindent $Q_{max} $ and $Q_{min}$ are the minimum and maximum values of the layer in the training phase. Therefore, the part $\frac{(2^{k}-1)}{Q_{max}-Q_{min}}$ could be calculated in advance, and this formula can be performed through in-memory addition and multiplication in subarrays.

Batch normalization is the following process that can recover the quantization loss and retain the accuracy of the model. The batch normalization transformation makes the data set have zero mean and one standard deviation \cite{ding2021improving}, and given below:

\begin{small}
\begin{equation}
I_{o} =\frac{I_{i}-\mu}{\sqrt{\sigma  ^{2}+\epsilon }}\gamma +\beta ,
\label{batch}
\end{equation}
\end{small}

\noindent where $I_{o}$ and $I_{i}$ denote the corresponding output and input of the transformation, respectively. $\sigma $ and $\mu$ are two statistics of the training model, $\gamma $ and $\beta$ are trained parameters used to restore the representation power of the network, and $\epsilon $ is a constant added for numerical stability. The aforementioned parameters are calculated and stored in advance, so that the above formula can be parallel performed through in-memory addition and multiplication in subarrays, similar to quantization. In addition, the ReLU activation function is achieved by replacing any negative number with zero. The MSB of the input is read out first and used to determine whether to write zero.

\textbf{Pooling layer}: Average pooling and max/min pooling are the two main types of pooling layers. Average pooling computes the average of all input values inside a sliding window. We support average pooling by summing the input values in a window and dividing the sum by the window size. Max/min pooling calculates the max/min of all the inputs inside the window and is accomplished by iterative in-memory comparison. In each iteration, the input for the comparison is selectively copied from max/min in the previous iteration.

\textbf{Fully-connected layer}: It has been concluded that the fully-connected layers can be implemented by convolution operations using 1$\times$1 kernels in networks \cite{ zhou2016dorefa, angizi2019mrima}. Therefore, we treat the fully-connected layer as convolutional layer.

\section{Evaluation}

\subsection{Platform Configurations}

To compare our design with other state-of-the-art solutions, we adopted a device-to-architecture evaluation along with an in-house simulator to evaluate the performance and energy benefits. We first characterized the hybrid circuit using a 45nm CMOS PDK and a compact Verilog-A model that is based on the Landau-Lifshitz-Gilbert equation \cite{wang2018high}. Table \ref{simu} lists some key device parameters used in our experiments. The circuit level simulation was implemented in Cadence Spectre and SPICE to obtain the performance parameters of basic logic operations. The results showed that it costs 180 fJ to erase an NAND-SPIN device with eight MTJs, with average 0.3 ns for each MTJ, and 840 fJ to program an NAND-SPIN device, with 5 ns for each bit. And the latency and energy consumption were 0.17 ns and 4.0 fJ for a read operation. The bit-counter module was designed based on Verilog HDL to obtain the number of non-zero elements. We synthesised the module with Design Compiler and conducted a post-synthesis simulation based on 45nm PDK. Secondly, we modified NVSim simulator \cite{7544313}, so that it calibrates with our design while performing access and in-memory logic operations. After configuring NVSim based on the previous results, the simulator reported the memory latency, energy and area corresponding to the PIM platform. Finally, for the architecture level simulation, we simulated the CNN inference tasks with an in-house developed C++ code, which simulates the data movement and in-memory computation in each layer. 

\begin{table}[t]
\footnotesize
\caption{Simulation parameters}
\label{simu}
\tabcolsep 9pt 
\begin{tabular*}{\textwidth}{ll|ll}
\toprule
Spin Hall angle & 0.3 &  Exchange bias & \qquad \qquad  15 mT \\
Gilbert damping & 0.02 &  TMR & \qquad \qquad  120$\%$ \\
Resistance-area product & 5 $\Omega\cdot\mu m^{2}$ & Tunneling spin polarization & \qquad \qquad  0.62\\
Saturation magnetization &  1150 kA/m & Heavy metal thickness & \qquad \qquad  4 nm \\
Ratio of damping-like SOT to field-like SOT & 0.4 & Uniaxial anisotropy constant & \qquad \qquad  1.16 $\times 10^{6} J/m^{3}$\\
\bottomrule
\end{tabular*}
\end{table}

\subsection{Experimental Setup}

\begin{figure}[t]
\centering
\includegraphics[width=0.7\textwidth]{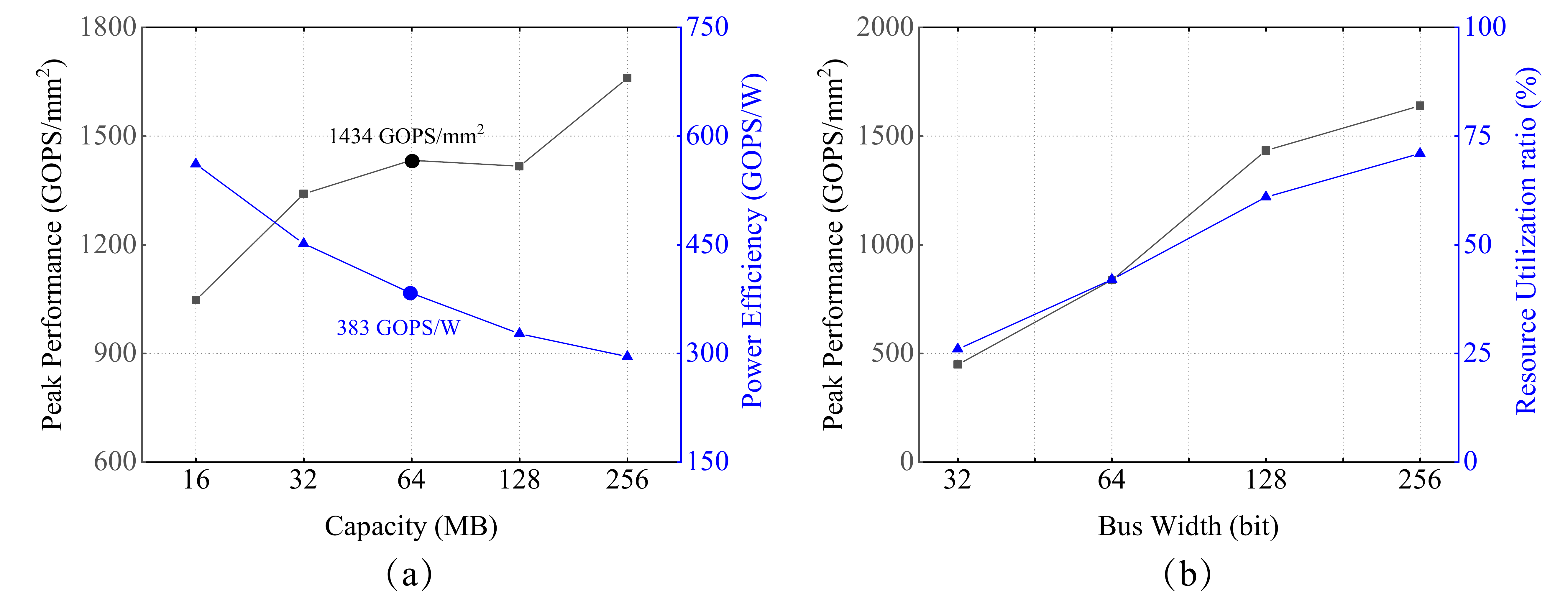}
\caption{(a) The effect of the capacity on the peak performance and energy efficiency. (b) The effect of the bus width on the peak performance and resource utilization ratios.}
\label{cap_bus_en}
\end{figure}

Both the memory capacity and bandwidth can affect the peak performance of the CNN accelerator. We evaluated these impacts on the basis of fixed memory structure. In our design, we assumed that there are 4$\times$4 subarrays with 256 rows and 128 columns in each mat, and 4$\times$4 mats were considered as a group. 

Obviously, enlarging the memory capacity brings a higher performance owing to the increase in the number of computation units. Fig.~\ref{cap_bus_en}a indicates the relationship between the performance and memory capacity. We observed that the peak performance normalized to the area tended to increase slowly with the expansion of the memory capacity, and it reached a regional peak at 64 MB. Nonetheless, the power efficiency dropped because of the increasing energy consumption of peripheral circuits.

Due to the bandwidth limitation, the architecture exhibited a relationship between the performance and the bandwidth as shown in Fig.~\ref{cap_bus_en}b. In addition, the weight data were transferred to subarrays through the bus and buffered in the buffer. Obviously, the peak performance normalized to the area rose linearly as the bandwidth increases. This mainly arises from that the higher bandwidth provided more data for computation units, which could also be verified from the view of hardware utilization ratios.

With reference to the above results, we configured our PIM architecture with a 64 MB memory array and a 128-bit bandwidth in subsequent simulations.

\subsection{CNN Acceleration Performance}
For comparison with state-of-the-art CNN accelerators, we regard the designs based on DRAM (DRISA in \cite{li2017drisa}), ReRAM  (PRIME in \cite{chi2016prime}), STT-RAM (STT-CiM in \cite{jain2017computing}, MRIMA in \cite{angizi2019mrima}), and SOT-RAM (IMCE in \cite{angizi2018imce}) as counterparts. Among various benchmarks, we validated the AlexNet/VGG19/ResNet50 models on the ImageNet dataset for a comprehensive evaluation. At runtime, the execution of convolution accelerators depends on the reasonable data flows and the control signals. The inputs and weights of each model were transferred to and initialized in subarrays. The complex logic operations in each layer were decomposed into a series of simple logic operations which were performed sequentially. Temporary results at runtime were transferred to each other across the buses between modules. Considering the uniqueness of those CNN models in depth and structure, the architectures had unique timing control signals to schedule the computations and communications for different models. In addition, the accelerators would split multi-bit data for fine-grained computations, when there was a mismatch between the data matrices and subarrays in size.

\begin{figure*}[tbp]
\centering
\begin{minipage}[t]{0.49\textwidth}
\setcaptionwidth{2.9in}
\centering
\includegraphics[width=0.9\textwidth,height=1\textwidth]{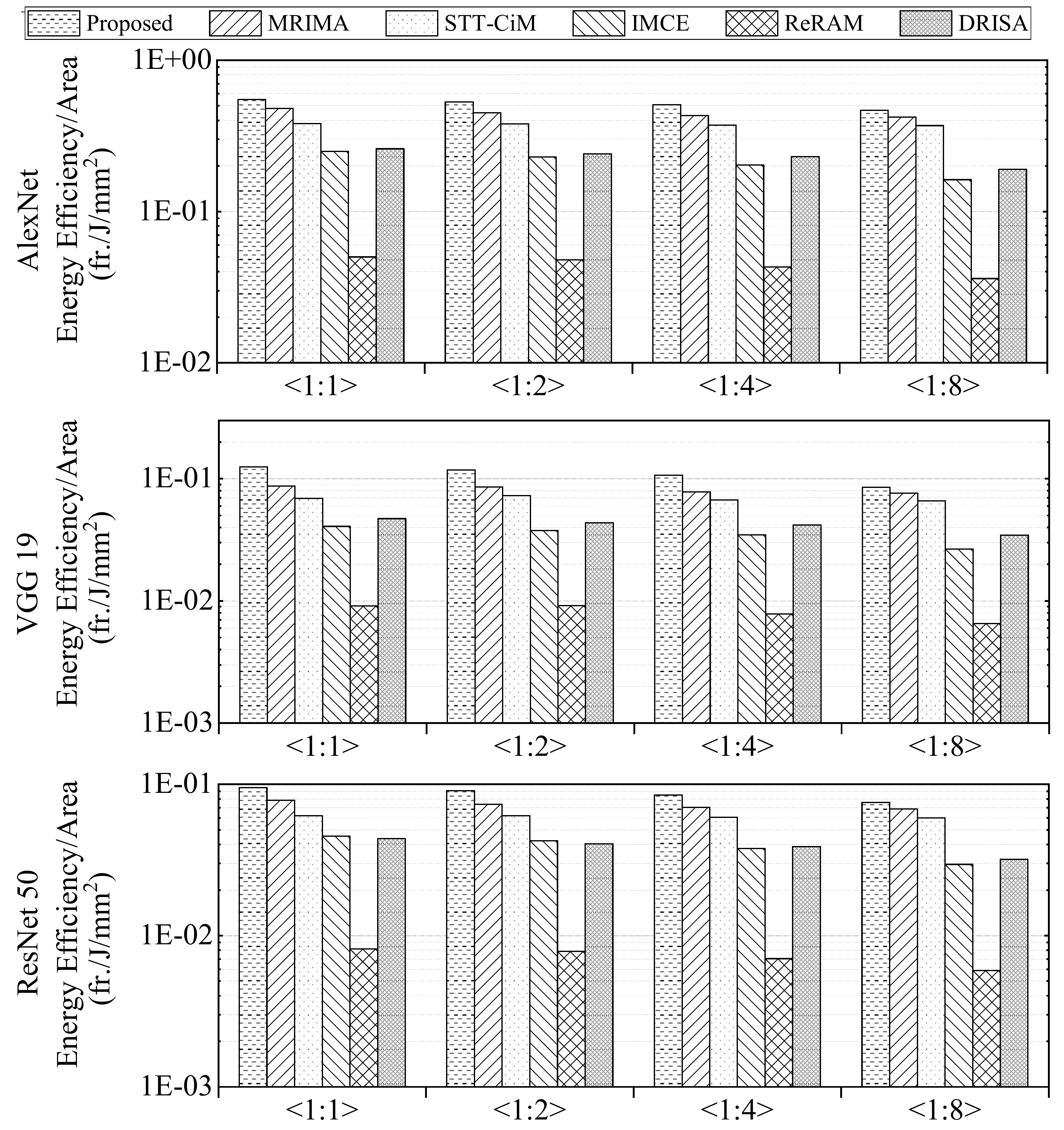}
\caption{ Comparison of the architecture efficiencies for different $\langle W:I \rangle$ ratios across various CNN models.}
\label{alex_v_r_e}
\end{minipage}
\begin{minipage}[t]{0.49\textwidth}
\setcaptionwidth{2.9in}
\centering
\includegraphics[width=0.9\textwidth,height=1\textwidth]{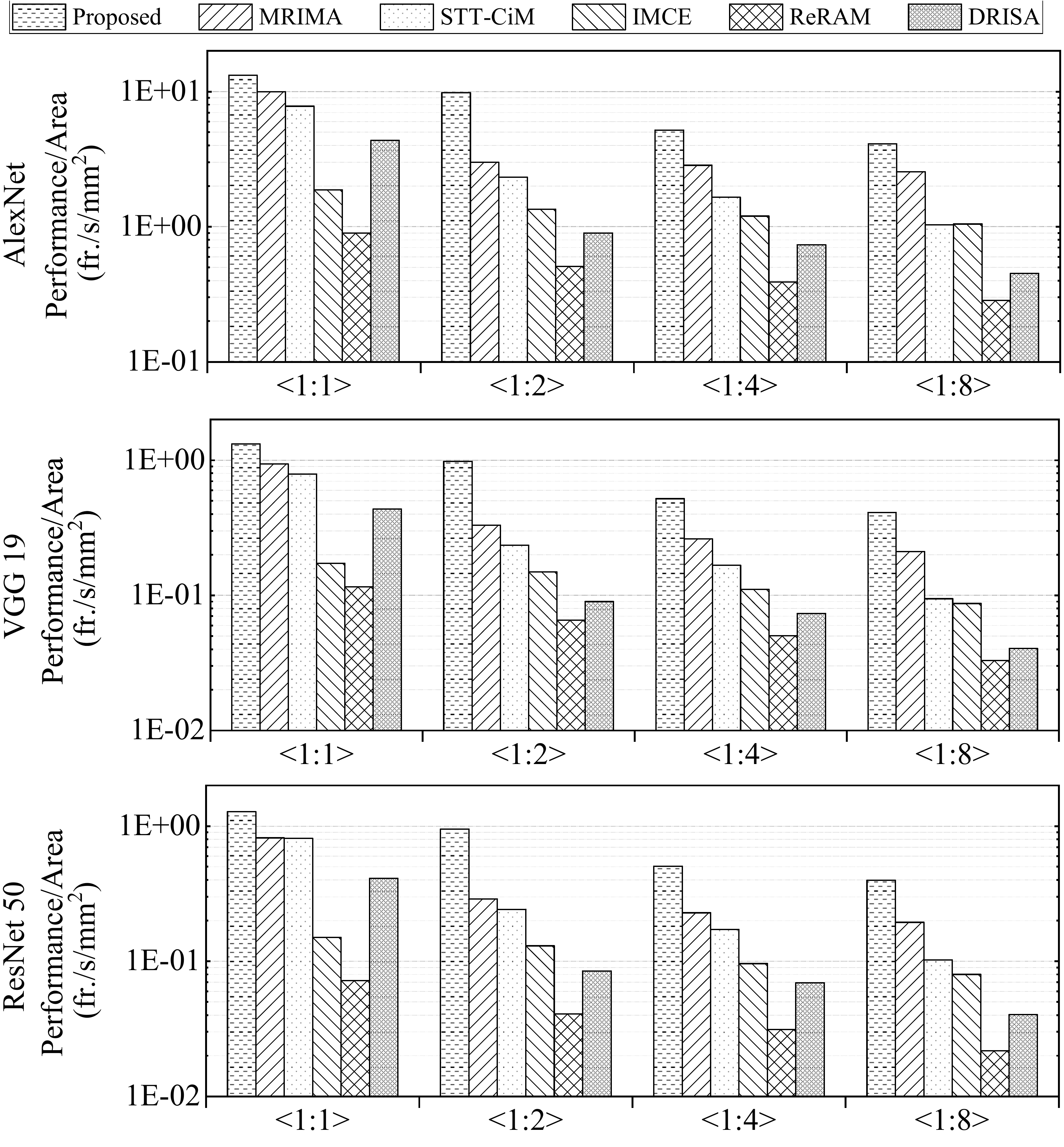}
\caption{ Comparison of the architecture performance for different $\langle W:I \rangle$ ratios across various CNN models.}
\label{alex_v_r_p}
\end{minipage}
\end{figure*}


\textbf{Energy efficiency}: We obtained the energy efficiency normalized to area results in different bit-width (precision) configurations $\langle W:I\rangle$ in three models. As shown in Fig.~\ref{alex_v_r_e}, our design offered energy efficiency superior to those of the other solutions. In particular, the proposed method achieved 2.3$\times$ and 12.3$\times$ higher energy efficiency than DRAM- and ReRAM-based accelerators on average, mainly for four reasons: 1) Part of the energy-intensive calculation was converted to efficient AND and bit-count operations. 2) The introduction of the buffer reduced data movements and rewrite operations within the memory, which increased the data reuse while reducing the energy consumption. This also contributed greatly to the superiority of our method to the SOT-based solution ($\sim$2.6$\times$ energy savings on average). 3) By exploiting the characteristics of the SOT mechanism and implementing the customized storage scheme, our architecture achieved lower energy consumption for data writing than all counterparts, even STT-CiM ($\sim$1.4$\times$ energy savings). 4) The elimination of complex functional units, such as ADCs/DACs in the ReRAM crossbar, also resulted in favorable energy efficiency. Although there were some adders and bit-counters in our design, the scheme in which different significant bits were separately processed dramatically reduces the number of accumulations. This is also why the improvement in the energy efficiency of our design becomes increasingly evident when $\langle W:I \rangle$ increases.

\textbf{Speedup}: The performance of each accelerator in different bit-width (precision) configurations $\langle W:I\rangle$ is presented in Fig.~\ref{alex_v_r_p}.  Among all solutions, our design obtained the highest performance normalized to area, with a 6.3$\times$ speedup over the DRAM-based solution and an approximately 13.5$\times$ speedup over the ReRAM accelerator. The improvement in our design comes from several aspects: 1) The parallel execution of logic operations and the pipeline mechanism for implementing accumulation fully utilized the hardware resources to complete efficient convolution calculation. 2) The participation of the buffer in PIM effectively reduced the in-memory data movements, data congestion, and bus competition, all of which reduce the waiting time. 3) There were no need for complex peripheral circuits in our design, such as ADCs/DACs in the ReRAM crossbar, which could reduce the area overhead to a certain extent. In addition, the results showed that our design is on average 2.6$\times$ and 5.1$\times$ faster than the STT-CiM and IMCE, mainly because of the efficient and parallel logic operations. 


\begin{table}[t]
\footnotesize
\caption{Comparison with related in-memory CNN accelerators}
\label{comp}
\tabcolsep 10pt 
\begin{tabular*}{\textwidth}{ccccccc}
\toprule
Accelerator & DRISA \cite{li2017drisa} & PRIME \cite{chi2016prime} & STT-CiM \cite{jain2017computing} & MRIME \cite{angizi2019mrima} & IMCE \cite{angizi2018imce} & Proposed \\
Technology &  DRAM & ReRAM & STT-RAM & STT-RAM  & SOT-RAM  & NAND-SPIN \\
Throughput (FPS) & 51.7 & 9.4 & 45.6& 52.3 & 21.8 & 80.6\\
Capacity (MB) &  64 & 64 & 64 & 64 & 64 & 64\\
Area ($mm^{2}$) & 117.2 & 78.2& 57.7& 55.6 & 128.3 & 64.5\\
\bottomrule
\end{tabular*}
\end{table}

Table \ref{comp} shows the area efficiency comparison of related in-memory CNN accelerators. We observed that STT-CiM and MRIMA show better area efficiency, which mainly comes from the high integration density of STT-MRAM-based memory designs. The SOT-MRAM-based architecture took the largest area, even more than the DRISA solution that embeds complex logic circuits in chips as the result of two transistors in a single cell. The proposed NAND-SPIN accelerator was not the most area-efficient architecture, but it offered the highest throughput by exploiting the data locality and benefiting from excellent characteristics of NAND-SPIN devices in memory arrays.

\begin{figure*}[tbp]
\centering
\begin{minipage}[t]{0.49\textwidth}
\setcaptionwidth{2.9in}
\centering
\includegraphics[width=1.0\textwidth]{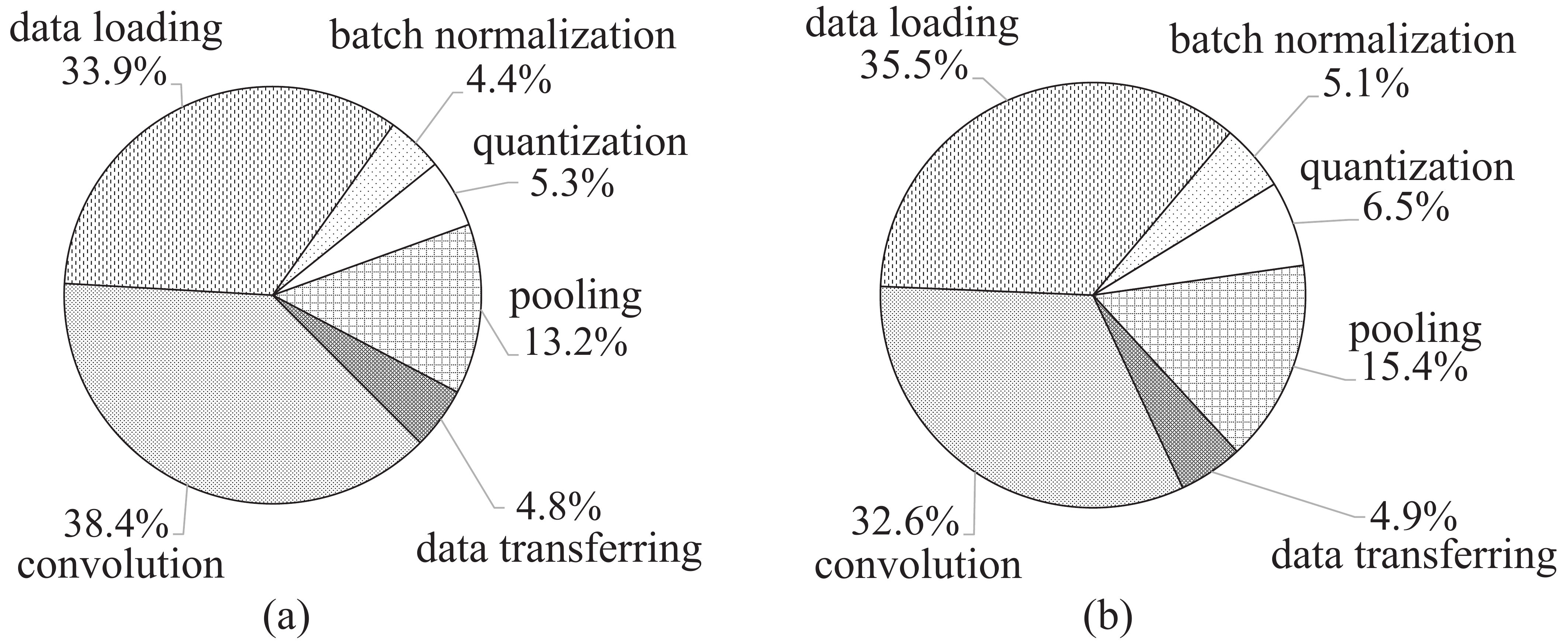}
\caption{ Breakdown of (a) latency and (b) energy.}
\label{breakdown}
\end{minipage}
\begin{minipage}[t]{0.49\textwidth}
\setcaptionwidth{2.9in}
\centering
\includegraphics[width=1.0\textwidth]{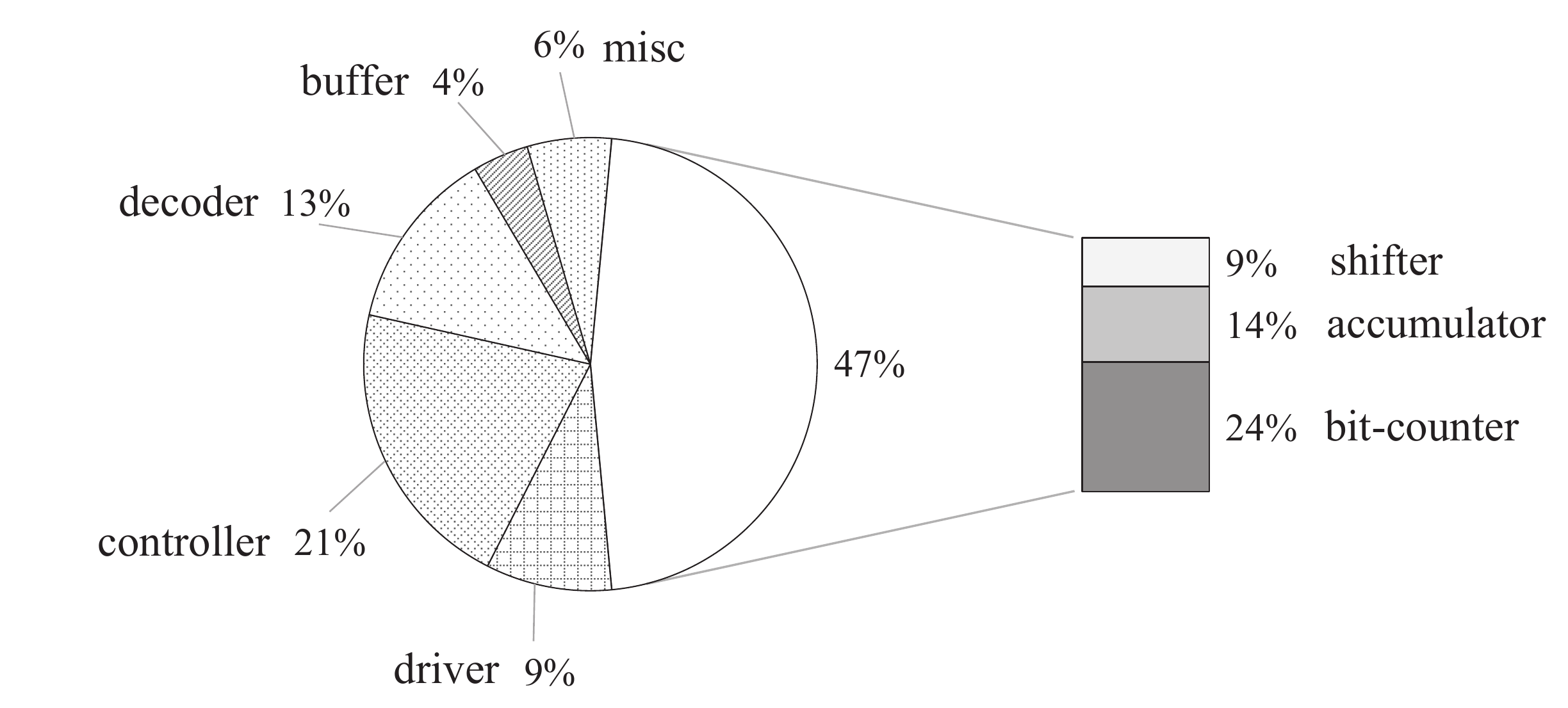}
\caption{ Area overhead breakdown.}
\label{area}
\end{minipage}
\end{figure*}

\textbf{Energy/Latency breakdown}: Fig.~\ref{breakdown} shows the latency and energy breakdown of our accelerator for ResNet50 model. In Fig.~\ref{breakdown}a, we observed that loading data and distributing them into arrays is the most time-consuming part, accounting for 38.4\%. This was mainly because writing data into NAND-SPIN device took more time than reading. The time spending on convolution and data transfer took 33.9\% and 4.8\% of the time respectively. In addition, 13.2\% of the time was spent on data comparison operations in the process of determining the maximum in pooling layers. The remaining parts were for batch normalization (4.4\%) and quantization (5.3\%).

As shown in Fig.~\ref{breakdown}b, the convolution, corresponding to numerous data reading and bit-counting operations, consumed the most energy up to 35.5\%. Due to the high writing energy consumption of NAND-SPIN device, loading data consumed nearly 32.6\% of the total energy consumption. Transferring data contributed to 4.9\% of the energy consumption, and 15.4\% of the energy was spent in pooling layers.  The other parts included batch normalization (5.1\%) and quantization (6.5\%).


\textbf{Area}: Our experiments showed that our design imposes 8.9\% area overhead on the memory array. The additional circuits supported the memory to implement in-memory logic operations and cache the temporary data in CNN computation. Fig.~\ref{area} shows the breakdown of area overhead resulted from the add-on hardware. We observed that up to 47\% area increase was taken by added computation units. In addition, approximately 4\% was the cost of the buffer, and other circuits, such as controllers and multiplexers, incurred 21\% area overhead. 

\section{Conclusion}

In this paper, we propose a memory architecture that employs NAND-SPIN devices as basic units. Benefiting from the excellent characteristics such as low write energy and high integration density, the NAND-SPIN-based memory achieves a fast access speed and large memory capacity. With supportive peripheral circuits, the memory array can work as either a normal memory or perform CNN computation. In addition, we adopted a straightforward data storage scheme so that the memory array reduces data movements and provides high parallelism for data processing. The proposed design exploits the advantages of PIM and NAND-SPIN to achieve high performance and energy efficiency during CNN inferences. Our simulation results demonstrate that the proposed accelerator can obtain on average $\sim$2.3$\times$ and $\sim$1.4$\times$ better energy efficiency, and $\sim$6.3$\times$ and $\sim$2.6$\times$ speedup than the DRAM-based and STT-based solutions, respectively.

\section*{Acknowledgement}

This work is supported in part by the National Natural Science Foundation of China (No. 62072019, 62004011, 62171013), the Joint Funds of the National Natural Science Foundation of China (No. U20A20204), and the State Key Laboratory of Computer Architecture (No. CARCH201917).







\end{document}